\documentstyle[12pt]{article}
\newcommand{\beq}{\begin{equation}}
\newcommand{\eeq}{\end{equation}}
\newcommand{\ba}{\begin{eqnarray}}
\newcommand{\ea}{\end{eqnarray}}

\newcommand{\dsl}
  {\kern.06em\hbox{\raise.15ex\hbox{$/$}\kern-.56em\hbox{$\partial$}}}

\newcommand{\eeqarr}{\end{eqnarray}}
\newcommand{\ZZ}{{\rm \kern 0.275em Z \kern -0.92em Z}\;}
\begin{document}
\begin{center}
{\Huge Duality in Noncommutative Topologically Massive Gauge Field 
Theory Revisited}
\\
\vspace*{0.7cm}
{\large M. Botta Cantcheff$^{*,\dag,}$\footnote{botta@cbpf.br, 
mbotta$\_$c@ictp.trieste.it} and
Pablo Minces$^{\dag,\ddag,}$\footnote{pablo@fma.if.usp.br}}
\\
\vspace*{0.5cm}
$^{*}$High Energy Section,\\
Abdus Salam ICTP, Strada Costiera 11, 34014 Trieste, Italy.\\
\vspace*{0.3cm}
$^{\dag}$Centro Brasileiro de Pesquisas F\'{\i}sicas (CBPF),\\
Departamento 
de Teoria de Campos e Part\'{\i}culas (DCP),\\ Rua Dr. Xavier Sigaud 150, 
22290-180, Rio de Janeiro, RJ, Brazil.\footnote{Permanent Address.}\\
\vspace*{0.3cm}
$^{\ddag}$Instituto de F\'{\i}sica Te\'orica (IFT), Universidade Estadual
Paulista (UNESP),\\
Rua Pamplona 145, 01405-900, S\~ao Paulo, SP, Brazil.
\end{center}
\begin{abstract}
We introduce a master action in noncommutative space, out of which we obtain
the action of the noncommutative Maxwell-Chern-Simons theory. Then, we look 
for the corresponding dual theory at both first and second orders in the 
noncommutative parameter. At the first order, the dual theory happens to be, 
precisely, the action obtained from the usual commutative Self-Dual model by 
generalizing the Chern-Simons term to its noncommutative version, including a 
cubic term. Since this resulting theory is also equivalent to the 
noncommutative massive Thirring model in the large fermion mass limit, we
remove, as a byproduct, the obstacles arising in the generalization to 
noncommutative space, and to the first nontrivial order in the noncommutative
parameter, of the bosonization in three dimensions. Then, performing 
calculations at the second order in the noncommutative parameter, we 
explicitly compute a new dual theory which differs from the noncommutative 
Self-Dual model, and further, differs also from other previous results, and 
involves a very simple expression in terms of ordinary fields. In addition, a 
remarkable feature of our results is that the dual theory is local, unlike 
what happens in the non-Abelian, but commutative case. We also conclude that 
the generalization to noncommutative space of bosonization in three dimensions
is possible only when considering the first non-trivial corrections over 
ordinary space.
\end{abstract}
\newpage

\section{Introduction}

The well-known duality between the Maxwell-Chern-Simons (MCS) theory
\cite{jackiw1} and the Self-Dual (SD) model \cite{townsend} was 
established long time ago in \cite{jackiw2}, both by comparing the 
corresponding equations of motion, and by introducing a master action 
out of which the MCS theory and the SD model can be 
obtained.\footnote{See \cite{lindstrom} for a review in the use of the 
master action approach in diverse areas, and \cite{ohta} for the 
application of the master action approach to the context of Bosonic 
p-branes.} Such duality 
leads 
to two equivalent descriptions of the 
dynamics of a parity violating, massive, spin one field. In particular, an 
important application of this duality is bosonization in three dimensions 
\cite{schaposnik1}\cite{banerjee2} of a theory of massive self-interacting 
fermions, namely, the massive Thirring 
(MT) model. Such bosonization was carried out in \cite{schaposnik1} 
by establishing, to leading order in the inverse fermionic mass, an 
identity between the partition functions of the MT and SD models, and 
then, by making use of the equivalence between the SD model and the MCS 
theory. In this way, the MT model is bosonized, and even when it has no 
manifest local gauge invariance, the bosonized theory is indeed a 
manifestly gauge invariant theory, namely, the MCS theory.

In view of the relevance of the above results, it would be interesting to
investigate the possibility of their extension to noncommutative (NC) 
space. During the last few years, the interest in NC 
field theories has intensified, due to applications in string theory 
\cite{douglas1}\cite{douglas2}\cite{witten1}, thus giving rise to a series 
of new developments and applications (see \cite{douglas3}\cite{szabo} for 
reviews). In NC space, the usual product is replaced by the star 
product of the form

\beq
{\hat g}(x)\star{\hat h}(x)=
exp\left[\frac{i}{2}\;\theta^{\alpha\beta}\partial^{{\hat g}}_{\alpha}
\partial^{{\hat h}}_{\beta}\right]{\hat g}(x){\hat h}(x) = {\hat 
g}(x){\hat h}(x) 
+\frac{i}{2}\;\theta^{\alpha\beta}\partial_{\alpha}{\hat g}(x)
\partial_{\beta}{\hat h}(x) + O(\theta^{2})\; ,
\label{1}
\eeq
where ${\hat g}(x)$ and ${\hat h}(x)$ are arbitrary functions, and the 
noncommutativity parameter $\theta^{\alpha\beta}$ is an antisymmetric 
constant tensor. A relation between NC and ordinary spaces is given by 
the Seiberg-Witten Map (SWM) \cite{witten1}, which interpolates between a 
NC gauge theory and its commutative counterpart, in such a way that NC 
gauge orbits are mapped into ordinary ones. The SWM on a NC Abelian gauge 
field ${\hat A}_{\mu}$ is given by

\beq
{\hat A}_{\mu} = A_{\mu} - 
\;\frac{1}{2}\;\theta^{\alpha\beta}
A_{\alpha}(2\partial_{\beta}A_{\mu}-\partial_{\mu}A_{\beta})+O(\theta^{2})\; 
.
\label{2}
\eeq
In particular, the NC-CS action of the form

\beq
I_{NC-CS}\sim \int d^{3}x\; \epsilon^{\alpha\mu\nu}{\hat 
A}_{\alpha}\star\left({\hat F}_{\mu\nu}+\;\frac{2i}{3}\;{\hat 
A}_{\mu}\star{\hat A}_{\nu}\right)\; , 
\label{3}
\eeq
was considered in \cite{grandi}, where it was shown that it 
is 
mapped, via Eq.(\ref{2}), into the standard commutative CS action. In the 
above equation, ${\hat F}_{\mu\nu}$ is
given by

\beq
{\hat F}_{\mu\nu}=\partial_{\mu}{\hat
A}_{\nu}-\partial_{\nu}{\hat A}_{\mu}-i\left[{\hat A}_{\mu} \;,\;
{\hat A}_{\nu}\right]_{\star}\; ,
\label{4}
\eeq 
where

\beq
\left[{\hat A}_{\mu}\;,\; {\hat A}_{\nu}\right]_{\star}={\hat
A}_{\mu}\star
{\hat A}_{\nu}-{\hat A}_{\nu}\star {\hat A}_{\mu}\; .
\label{5}
\eeq
Notice the presence in Eq.(\ref{3}) of a cubic term, even when we are 
considering the Abelian case. As shown in \cite{grandi}, such cubic term 
is cancelled out by the SWM. 

The extension to the NC case of the duality between MCS theory and 
the SD model was analyzed in \cite{ghosh1}\cite{ghosh2}\cite{dayi} 
(see \cite{our1} for results in the non-Abelian case). 
In particular, the proposal in \cite{ghosh1} was to consider calculations 
at the first non-trivial order in the NC parameter, and perform an inverse 
SWM  to the usual master action in commutative space, as defined in 
\cite{jackiw2}. This leads to a master 
action in NC space out of which to obtain, by following a procedure 
analogous to the one in \cite{jackiw2}, two theories 
which are in fact equivalent, and which could be 
considered, in principle, as the NC versions of the MCS theory and the SD 
model. In fact, the proposal in \cite{ghosh1} for the NC-MCS action is 
just the result obtained by performing the inverse SWM on the usual 
action of the MCS theory.

Later, Ref.\cite{ghosh2} considered the generalization to the NC plane of 
the bosonization of the MT model as performed in \cite{schaposnik1}. It 
was shown that the NC-MT model bosonizes, in the large fermion mass 
limit, into a model described, up to some conventional 
multiplicative coefficient, by the following action in NC space

\beq
I_{B}=\int d^{3}x\; \left[ -\mu {\hat f}^{\mu}\star {\hat f}_{\mu} 
+\;\frac{\kappa}{2}\;\epsilon^{\alpha\mu\nu}{\hat 
f}_{\alpha}\star\left(\partial_{\mu}{\hat f}_{\nu}-\;\frac{2i}{3}\;{\hat 
f}_{\mu}\star {\hat f}_{\nu}\right)\right]\; ,
\label{6}
\eeq
where $\mu$ and $\kappa$ are constant coefficients. Since the above action 
differs from the version of the NC-SD model given in \cite{ghosh1}, 
the conclusion in \cite{ghosh2} was that, even when the NC-MT model can be 
bosonized in powers of the inverse fermion mass, the duality between NC-MT 
model and NC-MCS theory is lost. It was stated that NC-MCS is not the dual 
of the NC-MT model.

On the other hand, the proposal in \cite{dayi} was to consider the NC-MCS 
theory as defined, up to some multiplicative coefficient, by the action 
(whose non-Abelian version was also considered in \cite{our1})

\beq
I_{NC-MCS}=\int d^{3}x\; \left[-\frac{\kappa ^{2}}{2\mu}\;{\hat 
F}^{\mu\nu}\star {\hat F}_{\mu\nu}-\kappa\epsilon^{\alpha\mu\nu}{\hat
A}_{\alpha}\star\left({\hat F}_{\mu\nu} + 
\frac{2i}{3}\;{\hat A}_{\mu}\star {\hat A}_{\nu}\right)\right]\; . 
\label{7}
\eeq
Our particular choice of the multiplicative coefficients will be 
clarified later.
It can be verified that, in fact, the above action differs from the 
one 
obtained in \cite{ghosh1} by performing an inverse SWM on the usual 
commutative action of the MCS theory. Notice, however, that Eq.(\ref{7}) 
is 
also a natural choice, as it makes use of the usual 
expressions of the NC Maxwell and CS theories. We emphasize that, even 
when 
the NC-CS action Eq.(\ref{3}) is mapped into its commutative version via 
the SWM, the 
same does not happen to the Maxwell term. 

The formulation in \cite{dayi}, which considers calculations at the second 
order 
in the NC parameter, involves to write Eq.(\ref{7}) in terms of 
ordinary fields, and then to define a master action 
out of which it 
can be obtained. Then, starting from such master action, Ref.\cite{dayi} 
computed another version of the NC-SD model, which differs from the one 
considered in 
\cite{ghosh1}\cite{ghosh2}, and involves an expression which is 
written in 
terms of commutative fields.

At this point we conclude that, in fact, there 
exists an 
ambiguity in the proper definition of the NC generalizations of the MCS 
theory 
and the 
SD model, and this fact was indeed recognized in Ref.\cite{ghosh2}, which 
wondered what theory should be referred to as the NC-SD 
model. One possible choice would be the one obtained by 
replacing the 
ordinary product by the star one, as in \cite{ghosh1}. Such version has 
the advantage that it obeys the self-dual equation, and is the one 
adopted in \cite{ghosh2}. On the other hand, 
Eq.(\ref{6}) is perhaps a more natural choice, as the mass term remains as 
such, and the CS term is generalized to its NC version Eq.(\ref{3}). In 
addition, as we have pointed out before, the action of the NC-MCS theory 
considered in \cite{ghosh1}\cite{ghosh2} differs from the one introduced 
in \cite{dayi}, namely Eq.(\ref{7}).

Summarizing, we are facing two difficulties in the generalization to the 
NC space of the duality between MCS theory and the SD model, namely, the 
existence of ambiguities when defining their corresponding NC 
versions, and the fact that a complete bosonization, along the lines of 
\cite{schaposnik1}, has not been formulated yet (even when 
Ref.\cite{ghosh2} managed to solve a part of the problem).

The purpose of this paper is to try to shed some light on the above 
detailed problems, by performing a careful analysis at a 
perturbative order in the NC parameter. 

A strong indication on the way to do this was given in \cite{our1}, 
dealing with non-Abelian CS theories in NC space, where it was shown, 
using the traditional master action approach at a 
perturbative level in the field but to all orders in the NC parameter, 
that the 
NC Yang-Mills-Chern-Simons action of the form

\beq
I_{NC-YMCS}=\int d^{3}x\; Tr\left[-\frac{\kappa ^{2}}{2\mu}\;{\hat{\cal
F}^{\mu\nu}}\star {\hat{\cal 
F}_{\mu\nu}}-\kappa\epsilon^{\alpha\mu\nu}{\hat{\cal
A}_{\alpha}}\star\left({\hat{\cal F}_{\mu\nu}} +
\frac{2i}{3}\;{\hat{\cal A}}_{\mu}\star {\hat{\cal 
A}}_{\nu}\right)\right]\; ,
\label{7'}
\eeq
where ${\hat{\cal A}}_{\mu}$ is a NC field in the adjoint representation 
of an 
arbitrary non-Abelian gauge group, is dual to an action which 
differs from the non-Abelian NC-SD model only by terms of the fourth order 
in the field, namely

\ba
I&=&I^{non-Abelian}_{NC-SD}+O\left({\hat {\cal 
B}}^{4}\right)\nonumber\\ &=&\int 
d^{3}x\; 
Tr\left[ -\mu {\hat {\cal B}}^{\mu}\star {\hat 
{\cal B}}_{\mu}
+\;\frac{\kappa}{2}\;\epsilon^{\alpha\mu\nu}{\hat 
{\cal B}}_{\alpha}\star\left(\partial_{\mu}{\hat 
{\cal B}}_{\nu}-\;\frac{2i}{3}\;{\hat  
{\cal B}}_{\mu}\star {\hat {\cal 
B}}_{\nu}\right)\right]+O\left({\hat {\cal B}}^{4}\right)\; .
\label{6'}
\ea
The master action proposed in \cite{our1} was the natural generalization 
of 
the one which is usually utilized in the commutative case 
\cite{jackiw2}\cite{schaposnik1}\cite{bota} (see also \cite{baner3} for a 
master action which has a gauge invariance in all fundamental fields). 

The above considerations motivate us to wonder what the results in 
the 
NC Abelian case would be like. We suspect that, since noncommutativity 
resembles
in some respects non-Abelian structures (notice, for example, the presence 
of a cubic term in the Abelian NC-CS action Eq.(3)), then it will be 
the case that, in the NC Abelian situation, the fourth order term in the
above equation is still present. However, such result arises when 
considering calculations involving all orders in the NC parameter. In 
principle, it could be possible that more useful results would arise when 
considering a perturbative approach. In that respect, we point out that, 
till date, most results in NC space involve corrections of the first order 
in $\theta$ over ordinary 
space, and this encourages us to perform calculations at orders 
$O(\theta)$ and $O(\theta^{2})$ and see what our results look like.

Going on this line of thought, our proposal here is to find, 
by performing calculations at a perturbative level, the dual 
theory of Eq.(\ref{7}), written in terms of ordinary fields. In doing 
this, we will separate our calculations into two stages. At the first 
part of calculations, performed in Section 2, we will find that, at order 
$O(\theta)$, the dual theory of 
$I_{NC-MCS}$ (as defined through 
Eq.(\ref{7})) is precisely Eq.(\ref{6}). Notice that this allows, by using 
the result in \cite{ghosh2} that Eq.(\ref{6}) is obtained by bosonizing 
the NC-MT model in the large fermion mass limit, to remove the obstacles 
arising in the generalization of the 
bosonization in three dimensions, along the lines of \cite{schaposnik1}, 
to the NC case. We emphasize that this result holds provided that only the 
first non-trivial corrections over ordinary space are considered.

Then, at the second part of our calculations, performed in Section 3, we 
will find that,
at order $O(\theta^{2})$, the 
`non-Abelian-like' nature of NC tehories finally prevails, and the duality 
between Eqs.(\ref{6}, \ref{7}) is lost.\footnote{The conjecture that 
duality should be lost when
considering higher orders of the NC generalization of the theory is also
suggested by the
recent result in \cite{miao}, which considers a NC chiral boson action and
shows that, for such model, self-duality is not maintained.} However, by 
computing the explicit form of the dual theory, we will show a remarkable 
result, namely, that it is local, unlike what happens to the non-Abelian, 
but commutative case \cite{karlhede}. In addition, we will show that our 
dual theory differs from the one computed in \cite{dayi}, and involves a 
much simpler expression written in terms of ordinary fields. We will also 
discuss the form of higher order contributions.

\section {First Order}

We begin our calculations by introducing the following master action in NC 
space

\beq
I_{M}=\int d^{3}x\; \left[-\mu {\hat f}^{\mu}\star {\hat 
f}_{\mu}+\kappa\epsilon^{\alpha\mu\nu}\left({\hat f}_{\alpha}\star 
{\hat F}_{\mu\nu} -{\hat   
A}_{\alpha}\star\left({\hat F}_{\mu\nu} +
\frac{2i}{3}\;{\hat A}_{\mu}\star {\hat A}_{\nu}\right)\right)\right]\; .
\label{8}
\eeq
In order to show the duality between the actions Eqs.(\ref{6}, 
\ref{7}), we will verify that solving $I_{M}$, first for ${\hat 
f}_{\mu}$ (in terms of ${\hat A}_{\mu}$) and then for ${\hat A}_{\mu}$ 
(in terms of ${\hat f}_{\mu}$), we recover both actions Eqs.(\ref{7},
\ref{6}), respectively. Throughout this paper, we consider boundary 
conditions such as surface 
terms in the action vanish. We begin by focusing on the NC-MCS theory 
Eq.(\ref{7}). From Eq.(\ref{8}), we find the following equation of 
motion for ${\hat f}^{\mu}$

\beq
{\hat f}^{\mu} = \frac{\kappa}{2\mu}\;\epsilon^{\mu\alpha\beta}{\hat 
F}_{\alpha\beta}\; ,
\label{9}
\eeq
and introducing this back into Eq.(\ref{8}), we get the NC-MCS action 
Eq.(7).

Now we focus on the NC-SD model Eq.(\ref{6}). We compute the equation of 
motion for 
${\hat A}_{\mu}$ in Eq.(\ref{8}). By performing calculations analogous to 
the ones in \cite{our1}, we arrive at 

\beq
2{\hat F}_{\mu\nu} = \partial_{\mu}{\hat f}_{\nu} - \partial_{\nu}{\hat 
f}_{\mu} -i[{\hat A}_{\mu},{\hat f}_{\nu}]_{*}+i[{\hat 
A}_{\nu},{\hat f}_{\mu}]_{*}\; .
\label{10}
\eeq
We first consider calculations at the first non-trivial order in $\theta$. 
In order to solve the above equation, we will write it in terms of 
ordinary fields $(A_{\mu},f_{\mu})$. This is done by 
performing a SWM 
of 
the form Eq.(\ref{2}) to the gauge field ${\hat A}_{\mu}$. In addition, 
and taking into account that, in a non-gauge theory, noncommutativity 
affects only products of fields in the action, without
changing the fields structures, we should also set the simple 
identity ${\hat f}_{\mu}=f_{\mu}$ \cite{ghosh1}. However, and just in 
order to be general, we will instead consider a mapping of the form

\beq
{\hat f}_{\mu}=f_{\mu}+\theta^{\alpha\beta}b_{\mu\alpha\beta}(f_{\nu})+ 
O(\theta^{2})\; ,
\label{10'}
\eeq
where $b_{\mu\alpha\beta}(f_{\nu})$ is an arbitrary function of $f_{\nu}$. 
In particular, the natural choice ${\hat f}_{\mu}=f_{\mu}$ corresponds to 
the particular case $b_{\mu\alpha\beta}=0$. We will show that, in fact, 
our final result (the duality, at the first non-trivial order in 
$\theta$, between the actions Eqs.(\ref{6}, \ref{7})) 
holds for any choice of $b_{\mu\alpha\beta}$. The only 
assumption that we will make is that $b_{\mu\alpha\beta}$ can be expanded 
as

\beq
b_{\mu\alpha\beta}(f_{\nu})=\sum_{n\geq 
0}b^{(n)}_{\mu\alpha\beta}(f_{\nu})\; ,
\label{10''}
\eeq
where $b^{(n)}_{\mu\alpha\beta}$ is of order $O(f^{n})$.

From Eqs.(\ref{1}, \ref{2}, \ref{4}, \ref{10'}), we 
write Eq.(\ref{10}) as

\ba
2F_{\mu\nu} + 
2\theta^{\alpha\beta}\left(F_{\mu\alpha}F_{\nu\beta}-A_{\alpha}\partial_{\beta}
F_{\mu\nu}\right)&=&\partial_{\mu}f_{\nu}-\partial_{\nu}f_{\mu} 
+\theta^{\alpha\beta}\left(\partial_{\alpha}A_{\mu}\partial_{\beta}f_{\nu}+
\partial_{\alpha}f_{\mu}\partial_{\beta}A_{\nu}\right)\nonumber\\ 
&+&\theta^{\alpha\beta}\left(\partial_{\mu}b_{\nu\alpha\beta}
-\partial_{\nu}b_{\mu\alpha\beta}\right)+ O(\theta^{2})\; ,
\label{11}
\ea
where $F_{\mu\nu}=\partial_{\mu}A_{\nu}-\partial_{\nu}A_{\mu}$. Now we 
must look for a solution $A_{\mu}[f_{\nu}]$ to the above equation. 
In order to do this, we follow \cite{our1}\cite{bota} and consider that 
the solution can be expanded as

\beq
A_{\mu}=\sum_{n\geq 0}A_{\mu}^{(n)}[f_{\nu}]\; ,
\label{12}
\eeq
where $A_{\mu}^{(n)}[f_{\nu}]$ is of order $O(f^{n})$. Thus, we will solve 
Eq.(\ref{11}) order by order in $f$, and also in $\theta$, by expanding 
each term $A_{\mu}^{(n)}$ in orders of $\theta$ as follows

\beq
A_{\mu}^{(n)}=\; ^{(0)}A_{\mu}^{(n)}\; + \;^{(1)}A_{\mu}^{(n)}(\theta)
+O(\theta^{2})\; .
\label{12'}
\eeq
In this way, $^{(p)}A_{\mu}^{(n)}$ is of order $n$ in $f$ and 
of order $p$ in $\theta$ (where $p=0,1$).

In solving Eq.(\ref{11}), we will drop 
pure-gauge 
terms of the form $^{(p)}F^{(n)}_{\mu\nu}=0$, due to the gauge 
invariance of Eq.(\ref{8}) (notice that the formulation in terms 
of ordinary 
fields inherits, via the SWM, an ordinary gauge invariance from the NC 
gauge invariance of Eq.(\ref{8})).

To the lowest order in $f$, we get from Eq.(\ref{11})

\beq
2F^{(0)}_{\mu\nu} + 
2\theta^{\alpha\beta}\left(F^{(0)}_{\mu\alpha}
F^{(0)}_{\nu\beta}-A^{(0)}_{\alpha}\partial_{\beta}
F^{(0)}_{\mu\nu}\right)=\theta^{\alpha\beta}
\left(\partial_{\mu}b^{(0)}_{\nu\alpha\beta}
-\partial_{\nu}b^{(0)}_{\mu\alpha\beta}\right)+O(\theta^{2})\; 
,
\label{13}
\eeq
and using Eq.(\ref{12'}) (i.e. solving order by order in $\theta$) we 
find, up to some pure-gauge term

\beq
A^{(0)}_{\mu}=\frac{1}{2}\;\theta^{\alpha\beta}
b^{(0)}_{\mu\alpha\beta}
+O(\theta^{2})\; .
\label{14}
\eeq

Next, to order $O(f)$ Eq.(\ref{11}) reads (notice that 
$^{(0)}A^{(0)}_{\mu}$ 
does 
not contribute)

\beq
2F^{(1)}_{\mu\nu}=\partial_{\mu}f_{\nu}-\partial_{\nu}f_{\mu}+
\theta^{\alpha\beta}\left(\partial_{\mu}b^{(1)}_{\nu\alpha\beta}
-\partial_{\nu}b^{(1)}_{\mu\alpha\beta}\right)
+O(\theta^{2})\; 
,
\label{16}
\eeq
which has solution (up to some pure-gauge term)

\beq
A^{(1)}_{\mu}=\frac{1}{2}\;f_{\mu}+ 
\frac{1}{2}\;\theta^{\alpha\beta}
b^{(1)}_{\mu\alpha\beta}(f_{\nu})+O(\theta^{2})\; .
\label{17}
\eeq

Now we consider the order $O(f^{2})$. From Eq.(\ref{11}) we get 
(we emphasize that $^{(0)}A^{(0)}_{\mu}$ does not contribute)

\ba
2F^{(2)}_{\mu\nu} +
2\theta^{\alpha\beta}\left(F^{(1)}_{\mu\alpha}F^{(1)}_{\nu\beta}
-A^{(1)}_{\alpha}\partial_{\beta}
F^{(1)}_{\mu\nu}\right)&=&
\theta^{\alpha\beta}\left(\partial_{\alpha}A^{(1)}_{\mu}
\partial_{\beta}f_{\nu}+
\partial_{\alpha}f_{\mu}\partial_{\beta}A^{(1)}_{\nu}\right)\nonumber\\ 
&+& 
\theta^{\alpha\beta}\left(\partial_{\mu}b^{(2)}_{\nu\alpha\beta}
-\partial_{\nu}b^{(2)}_{\mu\alpha\beta}\right)+O(\theta^{2})\; .
\ea
\label{18}
Using Eq.(\ref{17}), we find the following solution to the 
above equation (up to some pure-gauge term)

\beq
A^{(2)}_{\mu}=\theta^{\alpha\beta}\left[\frac{1}{8}\;f_{\alpha}
\left(2\partial_{\beta}f_{\mu}-\partial_{\mu}f_{\beta}\right)+
\frac{1}{2}\;b^{(2)}_{\mu\alpha\beta}(f_{\nu})+
H_{\mu\alpha\beta}(f_{\nu})\right] + O(\theta^{2})\; ,
\label{19}
\eeq
where
\beq
\partial_{\mu}H_{\nu\alpha\beta}=\frac{1}{8}\;\partial_{\alpha}f_{\mu}
\partial_{\beta}f_{\nu}\; .
\label{20}
\eeq

Finally, it can be shown that the solution to order $O(f^{n})$, with 
$n\geq 3$, is given by (up to some pure-gauge term)

\beq
A^{(n)}_{\mu}=\frac{1}{2}\;\theta^{\alpha\beta}
b^{(n)}_{\mu\alpha\beta}(f_{\nu})+ O(\theta^{2})\; \qquad\qquad (n\geq 3).
\label{20'}
\eeq

Summarizing, from Eqs.(\ref{10''}, \ref{14}, \ref{17}, \ref{19}, 
\ref{20'}) we get 
the following 
solution to Eq.(\ref{11})

\beq
A_{\mu}=\frac{1}{2}\;f_{\mu}+
\theta^{\alpha\beta}\left[\frac{1}{8}\;f_{\alpha}
\left(2\partial_{\beta}f_{\mu}-\partial_{\mu}f_{\beta}\right)+\frac{1}{2}
\;b_{\mu\alpha\beta}(f_{\nu})+
H_{\mu\alpha\beta}(f_{\nu})\right] + O(\theta^{2})\; ,
\label{21}
\eeq
where $H_{\mu\alpha\beta}(f_{\nu})$ satisfies Eq.(\ref{20}). 
From Eqs.(\ref{2}, \ref{21}) we find

\beq
{\hat 
A}_{\mu}=\frac{1}{2}\;f_{\mu}+\frac{1}{2}\;\theta^{\alpha\beta}
b_{\mu\alpha\beta}(f_{\nu})+\theta^{\alpha\beta}H_{\mu\alpha\beta}(f_{\nu})+ 
O(\theta^{2})\; .
\label{22}
\eeq
Now, introducing Eqs.(\ref{10'}, \ref{22}) into 
Eq.(\ref{8}), integrating by parts and using Eq.(\ref{20}), we arrive 
at our key result

\ba
I_{M}&=&\int d^{3}x\; [ -\mu 
f^{\mu}\left(f_{\mu}+2\theta^{\rho\beta}b_{\mu\rho\beta}\right)
+\;\frac{\kappa}{2}\;\epsilon^{\alpha\mu\nu} 
\left(f_{\alpha}+2\theta^{\rho\beta}b_{\alpha\rho\beta}\right)
\partial_{\mu}f_{\nu}\nonumber\\  &&\qquad\quad +\;
\frac{\kappa}{6}\;\epsilon^{\alpha\mu\nu}
\theta^{\rho\beta}
f_{\alpha}\partial_{\rho}f_{\mu}\partial_{\beta}f_{\nu}]+ O(\theta^{2})\; ,
\label{23} 
\ea
which, as can be verified using Eqs.(\ref{1}, \ref{10'}), corresponds 
precisely to the expansion, to the first 
non-trivial order in $\theta$, of Eq.(6). In this way, we have 
shown, at order O($\theta$), the equivalence between the 
theories described by the actions Eqs.(\ref{6}, \ref{7}), and used this 
result, together with the ones in \cite{ghosh2}, to remove the 
obstacles arising in the generalization to the NC space of the 
bosonization in three dimensions, along the lines of \cite{schaposnik1}. 
All calculations have been performed using the traditional master action 
approach, and considering only the first non-trivial order in 
the NC parameter. As emphasized before, this is not a very restrictive 
imposition, as most results computed till date involve corrections of 
order $O(\theta)$ over ordinary space.

\section{Second Order}

Now we analyze how our results extend to order $O(\theta^{2})$. For the 
sake of 
simplicity, we consider, instead of Eq.(\ref{10'}), the natural choice

\beq
{\hat f}_{\mu}=f_{\mu}\; .
\label{24}
\eeq 
We then note from Eq.(\ref{22}) that ${\hat A}_{\mu}$ will be of the form

\beq
{\hat
A}_{\mu}=\frac{1}{2}\;f_{\mu}+\theta^{\alpha\beta}
H_{\mu\alpha\beta}(f_{\nu})+\theta^{\alpha\beta}\theta^{\rho\sigma}
W_{\mu\alpha\beta\rho\sigma}(f_{\nu})+O(\theta^{3})\; ,
\label{25}
\eeq
where $H_{\mu\alpha\beta}(f_{\nu})$ satisfies Eq.(\ref{20}). In 
principle, 
$W_{\mu\alpha\beta\rho\sigma}(f_{\nu})$ should be computed by expanding
Eq.(\ref{10}) to order $O(\theta^{2})$ and then solving it. In 
order to do this, we should include the SWM to order $O(\theta^{2})$ in 
our calculations. 
However, nothing of this 
will be necessary, because an interesting result that we will show is 
that, in fact, $W_{\mu\alpha\beta\rho\sigma}(f_{\nu})$ will not contribute 
to our final result. 

Now using Eqs.(\ref{20}, \ref{24}, \ref{25}) we find

\ba
\epsilon^{\alpha\mu\nu}{\hat f}_{\alpha}\partial_{\mu}{\hat
A}_{\nu}&=&\frac{1}{2}\;\epsilon^{\alpha\mu\nu}f_{\alpha}\left(
\partial_{\mu}f_{\nu}+\frac{1}{4}\;\theta^{\rho\beta}\partial_{\rho}f_{\mu}
\partial_{\beta}f_{\nu}+2\theta^{\rho\beta}\theta^{\sigma\varphi}
\partial_{\mu}W_{\nu\rho\beta\sigma\varphi}(f)\right)\nonumber
\\&+&O(\theta^{3})\; ,
\label{26}
\ea

\ba
\epsilon^{\alpha\mu\nu}{\hat A}_{\alpha}\partial_{\mu}{\hat
A}_{\nu}&=&\frac{1}{4}\;\epsilon^{\alpha\mu\nu}f_{\alpha}(
\partial_{\mu}f_{\nu}+\frac{1}{2}\;
\theta^{\rho\beta}\partial_{\rho}f_{\mu}\partial_{\beta}f_{\nu}
+ 4\theta^{\rho\beta}\theta^{\sigma\varphi}\partial_{\mu} 
W_{\nu\rho\beta\sigma\varphi}(f)\quad\nonumber\\ &&\qquad\qquad
-\frac{1}{16}\;\theta^{\rho\beta}
\theta^{\sigma\varphi}\partial_{\rho}f_{\mu}
\partial_{\varphi}f_{\nu}\partial_{\beta}f_{\sigma})+O(\theta^{3})+\cdots\; 
,
\label{27}
\ea

\ba
\epsilon^{\alpha\mu\nu}{\hat f}_{\alpha}\left({\hat 
A}_{\mu}\star{\hat 
A}_{\nu}\right)&=&\frac{i}{8}\;\epsilon^{\alpha\mu\nu}f_{\alpha}
\left(\theta^{\rho\beta}\partial_{\rho}f_{\mu}\partial_{\beta}f_{\nu}-
\frac{1}{2}\;\theta^{\rho\beta}\theta^{\sigma\varphi}\partial_{\rho}f_{\mu}
\partial_{\varphi}f_{\nu}\partial_{\beta}f_{\sigma}\right)
\nonumber\\ &+& O(\theta^{3})\; ,
\label{28} 
\ea

\ba
\epsilon^{\alpha\mu\nu}{\hat A}_{\alpha}\left({\hat A}_{\mu}\star{\hat
A}_{\nu}\right)&=&\frac{i}{16}\;\epsilon^{\alpha\mu\nu}f_{\alpha}\left(
\theta^{\rho\beta}\partial_{\rho}f_{\mu}\partial_{\beta}f_{\nu}-
\frac{3}{4}\;\theta^{\rho\beta}\theta^{\sigma\varphi}\partial_{\rho}f_{\mu}
\partial_{\varphi}f_{\nu}\partial_{\beta}f_{\sigma}
\right)\nonumber\\&&
+O(\theta^{3})+\cdots\; ,
\label{29}
\ea
where the dots stand for total derivatives. Introducing 
Eqs.(\ref{26})-(\ref{29}) into Eq.(\ref{8}), the terms containing 
$W_{\nu\rho\beta\sigma\varphi}(f)$ remarkably cancel out, and we get

\ba
I_{M}&=&\int d^{3}x\; f_{\alpha}\left[ -\mu
f^{\alpha}+\frac{\kappa}{2}\;\epsilon^{\alpha\mu\nu}
\left(\partial_{\mu}f_{\nu}+\frac{1}{3}\;
\theta^{\rho\beta}
\partial_{\rho}f_{\mu}\partial_{\beta}f_{\nu}-\frac{1}{16}\;
\theta^{\rho\beta}\theta^{\sigma\varphi}\partial_{\rho}f_{\mu}
\partial_{\beta}f_{\sigma}\partial_{\varphi}f_{\nu}\right)\right]\nonumber\\ 
&+& O(\theta^{3})\; .
\label{30}
\ea
On the other hand, using Eq.(\ref{24}), the expansion to order 
$O(\theta^{2})$ of Eq.(\ref{6}) is given by

\ba
I_{B}&=&\int d^{3}x\; f_{\alpha}\left[ -\mu
f^{\alpha}+\frac{\kappa}{2}\;\epsilon^{\alpha\mu\nu}
\left(\partial_{\mu}f_{\nu}+\frac{1}{3}\;
\theta^{\rho\beta}
\partial_{\rho}f_{\mu}\partial_{\beta}f_{\nu}\right)\right]
+ O(\theta^{3})\; ,
\label{31}
\ea
so that

\beq
I_{M}=I_{B}-\frac{\kappa}{32}\;\epsilon^{\alpha\mu\nu}
\theta^{\rho\beta}\theta^{\sigma\varphi}
\int d^{3}x\; f_{\alpha}\partial_{\rho}f_{\mu}
\partial_{\beta}f_{\sigma}\partial_{\varphi}f_{\nu}+ O(\theta^{3})\; .
\label{32}
\eeq

In this way, we have shown that, as anticipated, the `non-Abelian-like' 
nature of NC theories finally prevails at order $O(\theta^{2})$, as the 
duality between Eqs.(\ref{6}, \ref{7}) is lost, due to the $f$-quartic 
term which has finally appeared at this order. However, a remarkable 
thing to notice is that the theory Eq.(\ref{30}) is local, unlike what 
happens to the non-Abelian, but commutative case \cite{karlhede}. Notice, 
also, that the Abelian nature of the theory allowed to compute the 
explicit expression of the $f$-quartic term, which is something that 
could not be done in the non-Abelian case (see Eq.(\ref{6'})). 

We should also contrast our results with the other previous ones in 
\cite{dayi}. Notice that, according to our calculations, the
dual of Eq.(\ref{7}) is given by Eq.(\ref{30}), which differs from
Eq.(17) of Ref.\cite{dayi}, and involves a much simpler 
expression. The 
remarkably simple form of Eq.(\ref{30}) followed from the key observation 
that $W_{\mu\alpha\beta\rho\sigma}(f_{\nu})$ in Eq.(\ref{25}) does not 
contribute to the final result.

In this work, we have considered calculations at orders $O(\theta)$ and 
$O(\theta^{2})$. In principle, this perturbative approach could be 
extended to higher orders in $\theta$. Notice that, at order 
$O(\theta^{3})$, the term $W_{\mu\alpha\beta\rho\sigma}(f_{\nu})$ in 
Eq.(\ref{25}) will finally contribute, and to compute its explicit form 
would involve 
to consider the correction at order $O(\theta^{2})$ over the SWM 
Eq.(\ref{2}). In principle, this poses no problem other that the 
increasing algebraic difficulties. However, it can be verified that, in 
solving Eq.(\ref{10}) 
at order $O(\theta^{3})$, new difficulties arise, which are similar to 
the ones already known from 
the non-Abelian, but commutative case, suggesting that, at order 
$O(\theta^{3})$, local solutions no longer exist. However, something 
that we were able to verify is that the $O(\theta^{3})$ correction over 
Eq.(\ref{30}) is of order $O(f^{5})$. This 
suggests that, in general, higher $O(\theta^{n})$ corrections over 
Eq.(\ref{30}) should be of order $O(f^{(n+2)})$.

To finish, we conclude by pointing out that three-dimensional 
bosonization in NC space should be possible only when NC 
effects are weak, and only the first non-trivial corrections over 
ordinary space, that is to say $O(\theta)$ corrections, are relevant. This 
is not a very restrictive imposition, as most results till date in NC 
spaces involve only such kind of corrections.

\section{Acknowledgments}

We would like to thank Prof. S. Ghosh for interesting comments. M.B.C is 
also grateful to Prof. F. A. Schaposnik for discussions. 
P.M would like to thank Prof. R. Jackiw for useful comments. M.B.C 
acknowledges financial support by CLAF. P.M. was supported by FAPESP grant 
01/05770-1 and by CLAF.

\end{document}